\documentclass[letterpaper,twocolumn,prl,aps,superscriptaddress,floatfix]{revtex4}
\usepackage[latin1]{inputenc}
\usepackage{bm}
\usepackage{amssymb,amsbsy,amsmath}
\usepackage{multirow,amssymb,amsbsy,amsmath}
\usepackage{graphicx}
\usepackage{verbatim}
\usepackage{color}
\usepackage{hyperref}
\usepackage{xcolor}
\usepackage{float}

\usepackage{amsmath}
\usepackage[normalem]{ulem}

\begin{document}

\title{Extensible universal photonic quantum computing with nonlinearity}

\author{Shang Yu\footnote{These authors contributed equally to this work}\footnote{shang.yu@imperial.ac.uk}}
\affiliation{Blackett Laboratory, Department of Physics, Imperial College London, London SW7 2AZ, United Kingdom}
\affiliation{Centre for Quantum Engineering, Science and Technology (QuEST), Imperial College London, London SW7 2AZ, United Kingdom}

\author{Jinzhao Sun$^{*}$\footnote{jinzhao.sun@qmul.ac.uk}}
\affiliation{School of Physical and Chemical Sciences, Queen Mary University of London, London E1 4NS, United Kingdom}

\author{Kuan-Cheng Chen$^{*}$}
\affiliation{Department of Electrical and Electronic Engineering, Imperial College London, London SW7 2AZ, United Kingdom}
\affiliation{Centre for Quantum Engineering, Science and Technology (QuEST), Imperial College London, London SW7 2AZ, United Kingdom}

\author{Zhi-Huai Yang}
\affiliation{HeliQ Standard CompuTech Co., Ltd, Hangzhou, 310015, China}

\author{Zhenghao Li}
\affiliation{Blackett Laboratory, Department of Physics, Imperial College London, London SW7 2AZ, United Kingdom}
\affiliation{Centre for Quantum Engineering, Science and Technology (QuEST), Imperial College London, London SW7 2AZ, United Kingdom}

\author{Ewan Mer}
\affiliation{Blackett Laboratory, Department of Physics, Imperial College London, London SW7 2AZ, United Kingdom}
\affiliation{Centre for Quantum Engineering, Science and Technology (QuEST), Imperial College London, London SW7 2AZ, United Kingdom}

\author{Yazeed K. Alwehaibi}
\affiliation{Blackett Laboratory, Department of Physics, Imperial College London, London SW7 2AZ, United Kingdom}
\affiliation{Centre for Quantum Engineering, Science and Technology (QuEST), Imperial College London, London SW7 2AZ, United Kingdom}

\author{Shana H. Winston}
\affiliation{Blackett Laboratory, Department of Physics, Imperial College London, London SW7 2AZ, United Kingdom}
\affiliation{Centre for Quantum Engineering, Science and Technology (QuEST), Imperial College London, London SW7 2AZ, United Kingdom}

\author{Dayne Marcus D. Lopena}
\affiliation{Blackett Laboratory, Department of Physics, Imperial College London, London SW7 2AZ, United Kingdom}
\affiliation{Centre for Quantum Engineering, Science and Technology (QuEST), Imperial College London, London SW7 2AZ, United Kingdom}

\author{Zi-Cheng Zhang}
\affiliation{Department of Physics, The University of Hong Kong, Hong Kong SAR, China}

\author{Guang Yang}
\affiliation{HeliQ Standard CompuTech Co., Ltd, Hangzhou, 310015, China}

\author{Runxia Tao}
\affiliation{College of Metrology Measurement and Instrument, China Jiliang University, Hangzhou 310018, China}
\affiliation{HeliQ Standard CompuTech Co., Ltd, Hangzhou, 310015, China}

\author{Mingti Zhou}
\affiliation{College of Metrology Measurement and Instrument, China Jiliang University, Hangzhou 310018, China}
\affiliation{HeliQ Standard CompuTech Co., Ltd, Hangzhou, 310015, China}

\author{Gerard J. Machado}
\affiliation{Blackett Laboratory, Department of Physics, Imperial College London, London SW7 2AZ, United Kingdom}
\affiliation{Centre for Quantum Engineering, Science and Technology (QuEST), Imperial College London, London SW7 2AZ, United Kingdom}

\author{Ying Dong\footnote{yingdong@cjlu.edu.cn}}
\affiliation{College of Metrology Measurement and Instrument, China Jiliang University, Hangzhou 310018, China}
\affiliation{HeliQ Standard CompuTech Co., Ltd, Hangzhou, 310015, China}

\author{Roberto Bondesan}
\affiliation{Department of Computing, Imperial College London, London SW7 2AZ, United Kingdom}
\affiliation{Centre for Quantum Engineering, Science and Technology (QuEST), Imperial College London, London SW7 2AZ, United Kingdom}

\author{Vlatko Vedral}
\affiliation{Clarendon Laboratory, University of Oxford, Parks Road, Oxford OX1 3PU, United Kingdom}

\author{M. S. Kim}
\affiliation{Blackett Laboratory, Department of Physics, Imperial College London, London SW7 2AZ, United Kingdom}
\affiliation{Centre for Quantum Engineering, Science and Technology (QuEST), Imperial College London, London SW7 2AZ, United Kingdom}

\author{Ian A. Walmsley}
\affiliation{Blackett Laboratory, Department of Physics, Imperial College London, London SW7 2AZ, United Kingdom}
\affiliation{Centre for Quantum Engineering, Science and Technology (QuEST), Imperial College London, London SW7 2AZ, United Kingdom}

\author{Raj B. Patel\footnote{raj.patel1@imperial.ac.uk}}
\affiliation{Blackett Laboratory, Department of Physics, Imperial College London, London SW7 2AZ, United Kingdom}
\affiliation{Centre for Quantum Engineering, Science and Technology (QuEST), Imperial College London, London SW7 2AZ, United Kingdom}

\begin{abstract}
Universal quantum computing requires an architecture that supports both linear circuits and, crucially, strong nonlinear resources. For quantum photonic systems, integrating such nonlinearities with scalable linear circuitry has been a major bottleneck, leaving most optical experiments without nonlinear operations and, consequently, incapable of achieving universality. Here, we report an extensible photonic computer that supports a universal gate set by seamlessly combining fully programmable, scalable linear optical networks with integrated nonlinear modules. This platform enables a broad range of quantum computing and simulation tasks. We demonstrate the quasi-deterministic generation of optical Gottesman-Kitaev-Preskill states, which are essential resources for bosonic error correction, yet had previously been realized only probabilistically. Furthermore, we simulate complex many-body quantum dynamics, exemplified by the Bose-Hubbard model. Such quantum simulation tasks have long been considered beyond the reach of photonic hardware limited to linear operations. These capabilities, enabled by our extensible architecture, establish a viable route towards photonic quantum simulation and fault-tolerant quantum computing.
\end{abstract}

\maketitle
\date{\today}

Quantum computing has recently achieved milestone demonstrations across various hardware platforms \cite{Rad2025,PsiQuantum2025,Larsen2025,Konno2024}, moving towards large-scale quantum computation beyond the noisy intermediate-scale quantum era. Photonic systems \cite{Bente2025,Obrien2009} continue to attract significant attention with advances ranging from architecture \cite{Rad2025,Bartolucci2023} and hardware \cite{PsiQuantum2025,Maring2024} to error correction \cite{Larsen2025,Konno2024} and applications \cite{Kawasaki2025,Lib2024,Yu2023}. While the long-term goal is to realize a large-scale universal photonic computer that incorporates a complete set of linear and nonlinear gates, recent experiments have separately demonstrated scalability \cite{Rad2025,PsiQuantum2025,Madsen2022}, programmability in linear optics \cite{PsiQuantum2025,Yu2023}, and access to nonlinear operations \cite{Biagi2022,Costanzo2017}.

Nevertheless, optical circuitry in these architectures remains non-universal \cite{Lloyd1999,Bartlett2002,Braunstein2005}, as single-photon-level nonlinearities are intrinsically difficult to realize and, even more challenging, to implement across all modes in large-scale linear photonic circuits \cite{Costanzo2017,Shapiro2006}. A flexible architecture capable of integrating diverse, function-specific photonic modules is essential to overcoming these challenges. Drawing inspiration from classical integrated circuit design, a natural approach is to develop an extensible and modular architecture that combines these components. {For instance, the specialized functional modules can be connected to a mainboard \cite{Rad2025} in a plug-and-play manner. However, this integration at the quantum level remains elusive; consequently, current fully programmable circuits are constrained to performing only linear operations \cite{PsiQuantum2025,Maring2024,Yu2023,Madsen2022}. An emerging approach instead is to exploit temporal multiplexing within a specially designed cyclic architecture. This method encodes information in a single spatial mode and allows the same physical modules to be reused across the circuit. It enables operations to be applied wherever needed and substantially reduces nonlinear-hardware costs in large-scale photonic circuits.}

\begin{figure*}[hbt]
\centering
\includegraphics[width=1.00\textwidth]{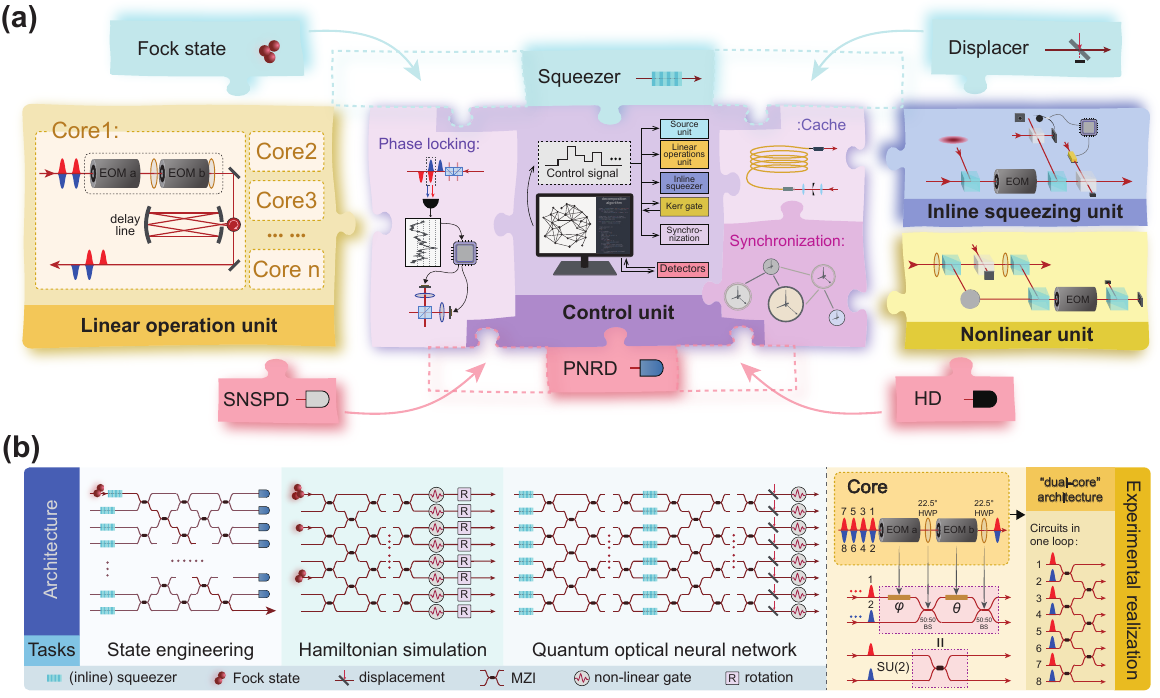}
\caption{\textbf{An extensible quantum photonic architecture and example applications.}
(\textbf{a}) Schematic of the architecture. \emph{Clavina} features jigsaw-like extensibility. A central control unit provides phase control, synchronization, and long-fiber delay lines that serve as a cache; the LOU, inline squeezing unit \cite{Miwa2014}, and nonlinear unit \cite{Costanzo2017} can be plugged into this control unit to expand the system's scale and functionality. Different types of light sources (blue pieces) and detectors (red pieces) can also be connected individually or simultaneously. This design achieves extensibility by combining mode-count scalability with modular composability.
(\textbf{b}) Several example applications offered by Clavina. Tasks such as sampling \cite{Yu2023,Madsen2022}, and cluster-state generation \cite{Larsen2019,Asavanant2019} can be directly implemented on its linear circuits. By integrating non-Gaussian resources, more advanced applications, for example, QONNs \cite{Yu2024QONN}, error-correction codes \cite{Konno2024,Winnel2024}, and Hamiltonian simulation \cite{Yang2020} also become achievable. Further details about these applications are provided in Supplementary Information Section VI. The diagram on the right provides a detailed illustration of how the ``cores'' enable interactions among qumodes.
}
\label{Fig1}
\end{figure*}

In this work, we present such an extensible all-optical platform \emph{Clavina}, which integrates a large-scale linear optical network with plug-and-play nonlinear modules via a central quantum photonic control unit. The fully connected, programmable linear network attains large scale via a multi-core time-bin interferometer. We further extend its functionality by inserting two addressable modules---an inline squeezer \cite{Miwa2014} and a Kerr gate \cite{Costanzo2017}, each of which can be independently enabled or disabled as required. Our design preserves the scalability and programmability of linear operations while incorporating nonlinear resources, and empowers the system to support a universal gate set \cite{Lloyd1999}. This achieves universal gate sets, and thus completes the final piece required for universal photonic computing. With this architecture, applications which were previously out of reach are unlocked. In this vein, we implement two applications that would otherwise be challenging to realize in existing experimental setups.

First, we demonstrate the preparation of large cluster states and non-Gaussian states. Unlike previous probabilistic schemes, Clavina enables the integration of a quantum light source that near-deterministically supplies photon-number-squeezed states, providing the requisite non-Gaussian resource. This advantage enables us to prepare small Schr\"{o}dinger cat states without post-selection, and convert them to larger amplitude cat states through two rounds of breeding \cite{Jeong2005}. Capitalizing on this, with real-time feed-forward, we achieve the quasi-deterministic generation of optical Gottesman-Kitaev-Preskill (GKP) states \cite{Gottesman2001}, a leading resource for bosonic quantum error correction.

\begin{figure*}[hbt]
\centering
\includegraphics[width=1.0\textwidth]{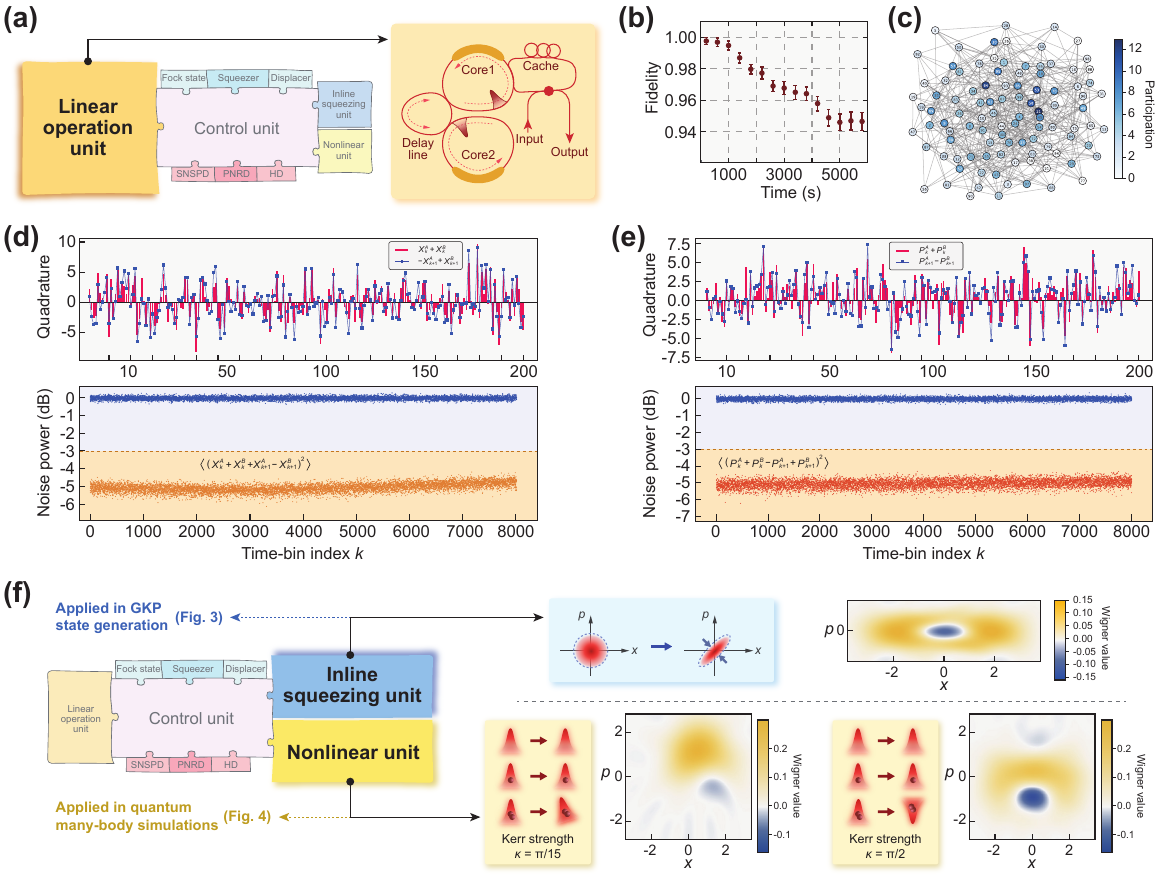}
\caption{\textbf{Scalability and nonlinearity through modular functional units.}
(\textbf{a}) Large-scale linear transformations using the LOU.
(\textbf{b}) By comparing sampling patterns acquired at different times with the initial pattern, we tracked the fidelity over nearly two hours, clearly demonstrating the platform's stability. Error bars denote the standard error. 
(\textbf{c}) Participation coefficients for all nodes in a 100-node network are extracted from Gaussian boson sampling data.
(\textbf{d} and \textbf{e}) Quadrature correlation of the cluster state indicates strong $x$- and $p$-quadrature correlations between modes $A$ and $B$ across the first 200 time-bin modes. The measurements below show nullifier variances consistently below the -3 dB inseparability threshold over 8,000 modes, confirming sustained multimode entanglement in both quadratures.
(\textbf{f}) Top: An inline squeezer acting on a single-photon state \cite{Miwa2014} could serve as a resource for generating GKP codes. Bottom: A Kerr gate \cite{Costanzo2017} with Hamiltonian $\hbar\kappa\hat{n}(\hat{n}-1)$ (where $\hat{n}$ is the bosonic number operator) directly introduces nonlinearity, demonstrated by the Wigner negativity (the blue regions). These plug-in units (including photon-number-state inputs) provide practical tools for achieving universality.}
\label{Fig2}
\end{figure*}

Secondly, Clavina can perform complex many-body quantum simulations, as exemplified by the Bose-Hubbard model \cite{Bloch2008},  beyond the constraints of linear optics. By connecting the programmable linear network to a tunable Kerr gate \cite{Costanzo2017} module, we can encode all the boson-lattice parameters. Combined with photon-number-resolving detectors (PNRDs), simulations of quantum dynamics beyond the hard-core boson limit (i.e., more than one boson per site) are possible, which is inherently challenging for superconducting platforms as their on-site interactions are non-tunable. These capabilities establish a new experimental paradigm for many-body simulation and reveal an all-optical pathway to photonic universal quantum simulation.

\vspace{6pt}
\noindent \textbf{Extensible photonic quantum processor}\\
The extensibility of Clavina is illustrated in Fig. 1(a) as a jigsaw assembled piece by piece. This extensible feature allows us to add new functionality to a circuit without a complete overhaul of the design. A central control unit---comprising phase-locking, cache (long-fiber delay-line), and synchronization modules---dispatches control signals to all active modules and schedules their tasks. By implementing time-bin encoding \cite{Yonezu2023,Madsen2022,Enomoto2021} within a loop-based design \cite{Yu2023}, the control unit docks reusable hardware along a single optical path to execute operations across sequential time bins, and additional functional units can be plugged into the control unit as needed to expand functionality. For example, the photon source unit---providing selectable squeezed or coherent light, or photon-number states via heralded detection---and the detection unit---offering superconducting nanowire single-photon detectors (SNSPDs), photon-number-resolving detectors (PNRDs), or homodyne detectors (HDs), as required---can be connected directly to the control unit's input and output ports. In addition, a linear operation unit (LOU) that supports multiple computation cores, shortens the optical path length for the same circuit depth and, together with phase-stabilized delay lines, implements large-scale linear transformations.

Notably, this architecture lets us complete the final piece required for universal photonic computing: a nonlinear unit incorporating a Kerr gate \cite{Costanzo2017} is coupled to the control unit, overcoming the limitations of purely linear operations. This satisfies the universality requirement in the continuous-variable setting, which demands access to both Gaussian and non-Gaussian resources \cite{Lloyd1999,Bartlett2002,Braunstein2005}. With this flexible and extensible architecture, additional functional modules can be integrated in a plug-and-play manner, making Clavina an upgradable photonic quantum computing system, and for a variety of applications, including quantum computing, simulation, and machine learning (Fig. 1(b)). For more details about the experimental setup and concrete applications, see Supplementary Information, Sections II-IV.

\begin{figure*}[hbt]
\centering
\includegraphics[width=0.99\textwidth]{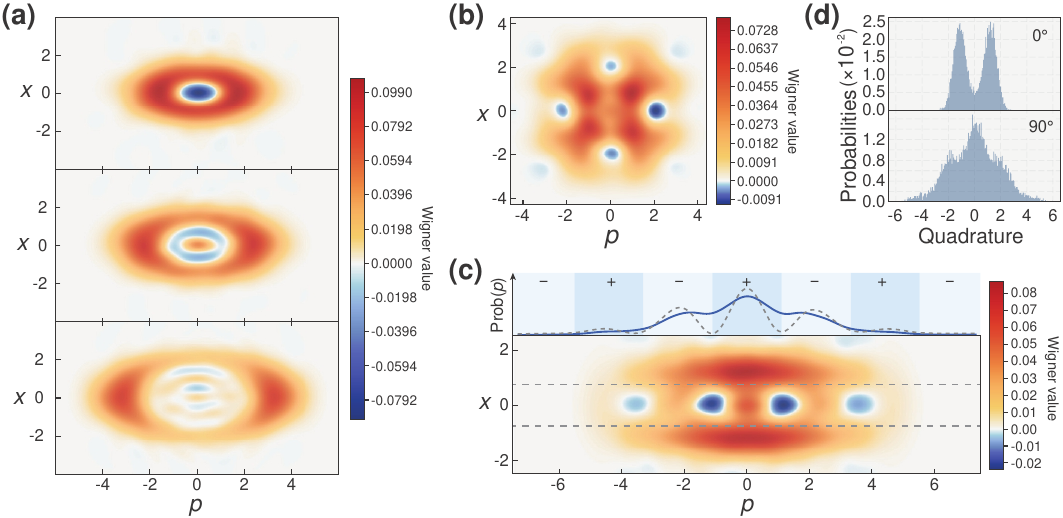}
\caption{\textbf{Non-Gaussian state engineering and quasi-deterministic GKP state generation.}
(\textbf{a}) Schr\"odinger cat state breeding outcomes. An initial small cat state is prepared by inline squeezing of a single-photon Fock state with squeezing parameter $r \approx 0.3$. It is then expanded into a larger-amplitude cat state through iterative breeding (middle: after one round; bottom: after two rounds).
(\textbf{b}) The Wigner function of the optical compass state is reconstructed from the interference of two cat states, prepared with inline squeezing $r=0.6$.
(\textbf{c}) Quasi-deterministic generation of the GKP grid state. The marginal probabilities, shown as a solid blue line for the full dataset and a gray dashed line for the subset filtered to $|x|<0.75$, reveal a phase-space lattice of peaks and troughs characteristic of a logical grid state. The initial cat state is prepared using inline squeezing with $r=0.48$.
(\textbf{d}) Measured quadrature probability distributions for the GKP state along two orthogonal phase quadratures (0 and $\pi/2$ cuts). More details are given in Supplementary Information Sec.VII B.}
\label{Fig3}
\end{figure*}

Clavina's scale and full programmability are demonstrated with a dual-core LOU (Fig. 2(a)), which maintains high stability as the system scales (Fig. 2(b)). We conducted a 100-mode Gaussian boson sampling experiment that requires adjustable squeezing and a fully reconfigurable interferometer to evaluate a network's participation coefficient \cite{Yu2023}, as shown in Fig. 2(c). In addition, Clavina readily time-multiplexes squeezed light pulses to generate scalable cluster states \cite{Asavanant2019,Larsen2019}. 
Fig. 2(d),(e) confirms the quantum correlations across 8,000 temporal modes. More information can be found in Supplementary Information Section VI.

Beyond linear operations, Clavina's extensibility offers two routes to introduce nonlinear resource required for universality (Fig. 2(f)). First, supply nonlinear resources at the input (e.g., photon-number states) together with a plug-in squeezing module. The top panel shows that an inline squeezer enables Gaussian processing while exploiting photon-number nonlinearity, thereby facilitating GKP code preparation. Second, insert a nonlinear module, such as a Kerr gate. The bottom panel shows that the Kerr gate applies a photon-number-dependent phase shift and, when acting on a coherent input, generates Wigner negativity, confirming successful nonlinearity injection. This flexibility in placing nonlinear operations within the circuit substantially advances the realization of universal photonic quantum computing.

\begin{figure*}[hbt]
\centering
\includegraphics[width=1.00\textwidth]{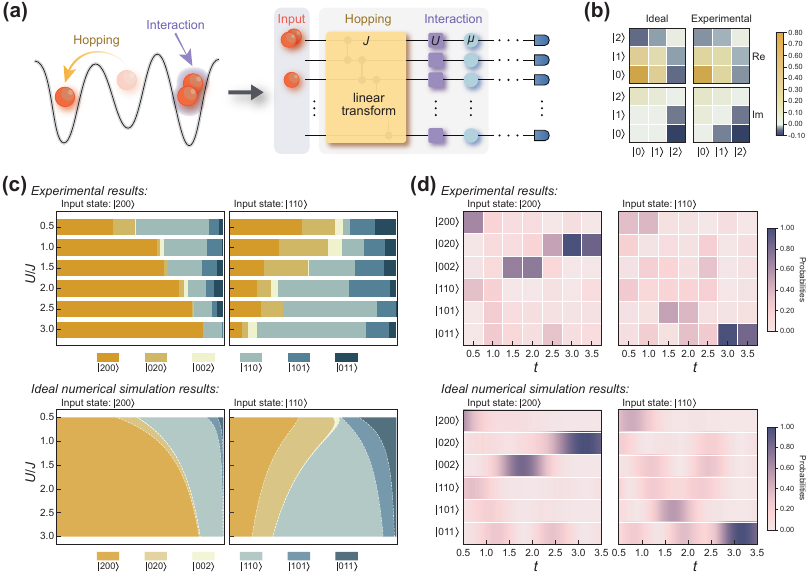}
\caption{\textbf{Nonlinear resource for simulating Bose-Hubbard Hamiltonian.}
(\textbf{a}) Kerr-gate fidelity on the low-photon Fock subspace. The corresponding fidelity reaches $0.820\pm 0.033$ compares with ideal value. Here, input is coherent state with $\alpha=0.5$, and the Kerr nonlinearity $\Phi=\pi/3$.
(\textbf{b}) A diagrammatic sketch of the one-dimensional Bose-Hubbard model shows bosons (e.g., photon) hopping between nearest-neighbor sites of a linear lattice while experiencing an on-site interaction. This model can be simulated with photonic quantum circuits, which can be realized on \emph{Clavina} with connecting Kerr module to main QPU. The hopping term is realized by the QPU, which couples time-bin modes in direct analogy to particle tunneling between neighboring lattice sites. The on-site interaction term, describing boson-boson collisions, is implemented via the Kerr module.
(\textbf{c}) Occupation probabilities after a fixed evolution time ($t=0.5$) while the interaction-to-tunneling ratio $U/J$ is swept from 0.5 to 3.0. Each bar is stacked according to the six two-photon Fock configurations. Left and right sub-panels correspond to the initial states $|200\rangle$ and $|110\rangle$, respectively. 
(\textbf{d}) Time resolved evolution at $U/J = 1$. Heat maps display the same set of Fock-state probabilities as a function of time for the two initial states used in Fig. 4(c). Top: Experimental results. Bottom: Ideal numerical simulations densely sampled across $U/J$ and $t$. The observed agreement, within experimental uncertainty, demonstrates that \emph{Clavina} accurately captures the dynamics.
}
\label{Fig4}
\end{figure*}

\vspace{6pt}
\noindent \textbf{Non-Gaussian state engineering}\\
Nonlinear resources are essential for non-Gaussian state generation and universal quantum computing, but are challenging to implement in photonic systems due to the absence of natural photon-photon interactions. Current demonstrations rely either on light-matter interactions \cite{Chang2014} or on measurement-induced schemes \cite{Patel2016,Brien2003}, but these approaches are typically weak \cite{Shapiro2006} or probabilistic, respectively.

Here, we employ a boosted \cite{Simon2024}, heralded photon-number-state generator to produce photon-number Fock states near-deterministically, thereby introducing nonlinear resources (see Supplementary Information Section II.C). We prepare these states at a 1 MHz repetition rate and inject them into an inline squeezing module (Fig. 1(b)) to generate odd Schr\"odinger cat states. The boosted heralding efficiency is approximately $93\%$. Taking residual circuit loss into account, the effective output rate is estimated to be $\approx 0.85$~MHz (see Supplementary Information for details). 
Enabled by the interferometric stability of Clavina, a subsequent two-round breeding process is performed in the LOU, in which interfering pairs of small cat states increase their amplitude to generate larger cat states \cite{Sychev2017}. Wigner tomography \cite{Lvovsky2009} results shown in Fig. 3(a) confirm both the growth in amplitude and the preservation of Wigner negativity. With a modified heralding pattern, i.e., conditioning on a two-photon detection event in the heralding PNRD arm, the same breeding protocol can prepare a compass state, a four-component coherent state superposition, which is shown in Fig. 3(b).

\vspace{6pt}
\noindent\textbf{Quasi-deterministic optical GKP state generation}\\
The cat states prepared above enable the synthesis of the optical GKP state \cite{Konno2024}. With Clavina, several cat states can interfere with each other in the LOU whilst monitoring each ancilla bin with one homodyne detector. The measurement outcomes of the ancilla time bins are fed-forward to a real-time displacement operation on the remaining time-bin. This allows a GKP state to be produced quasi-deterministically \cite{Weigand2018}---a key breakthrough toward practical quantum error correction. While recent probabilistic GKP state preparation schemes \cite{Konno2024, Larsen2025} have brought about significant advances, our architecture offers a deterministic approach, in principle, enabling the generation of cat states without post-selection and incorporating feedforward conditioned on homodyne outcomes \cite{Weigand2018}. In practice, we filter out homodyne outcomes with excessive noise---arising mainly from the limited throughput and squeezing available in the current implementation---which constrains the system to quasi-deterministic generation of GKP states. Here, by operating the experiment at 250~kHz, we generate approximately 2,000 GKP states per second, with a quality comparable to that shown in Fig. 3(c).

Fig. 3(c) shows the generated one-dimensional GKP state, with the quadrature distributions shown in Fig. 3(d) (see Supplementary Information Sec. VII B for more details). This reconstructed Wigner function exhibits the characteristic grid pattern, consistent with an approximate GKP logical state. We assess the quality of this GKP state by measuring the stabilizer amplitudes $\langle \hat{S}_{x}\rangle = 0.1061\pm0.0071$ and $\langle \hat{S}_{|1_{\text{L}}\rangle}\rangle = 0.2065\pm0.0055$. These results show that the state preserves appreciable lattice-wide phase coherence while suppressing the logical-$Z$ component, as expected for the $|1_{\text{L}}\rangle$ codeword. Independent confirmation comes from the quadrature variances of the central lattice peak, $\langle\Delta^{2}x\rangle = 0.1493\pm0.0052$ and $\langle\Delta^{2}p\rangle = 0.0870\pm0.0018$, whose product $\langle\Delta^{2}x\rangle\langle\Delta^{2}p\rangle = 0.0130\pm0.0005$ lies more than an order of magnitude below the vacuum limit, confirming the presence of sub-Planck-scale interference fringes characteristic of an ideal grid state~\cite{Zurek2001}. Further details such as stabilizer definitions, effective peak squeezing, are given in the Supplementary Information.

As with previous demonstrations\cite{Konno2024, Larsen2025}, lower loss and higher squeezing are required for these states to be practically useful for error correction. Despite this, the relatively large stabilizer amplitudes together with a sharply localized peak indicate that the state cannot be reproduced with purely Gaussian resources. Moreover, its quasi-deterministic generation from a near-on-demand non-Gaussian resource (i.e., Fock states) makes this approach a promising route to fault-tolerant quantum computing via additional Gaussian operations\cite{Baragiola2019}.

\vspace{6pt}
\noindent \textbf{Photonic quantum many-body simulations}\\
In the above studies, the nonlinear resource is provided by a squeezed-Fock-state. We now demonstrate an alternative, more direct approach with nonlinear gates. We integrate a Kerr gate---realized by interferometrically combining photon-addition and -subtraction sequences \cite{Costanzo2017,Biagi2022} (see Supplementary Information Section V)---into the linear circuits to induce strong nonlinearity. This Kerr module is shown in the lower-right panel of Fig. 1(a), and its fidelity is quantified in Fig. 4(b). Owing to the extensible design of Clavina, this module can be easily inserted into control unit, providing the essential non-Gaussian element that, together with our Gaussian toolbox, completes the universal gate set for photonic quantum computing \cite{Lloyd1999,Bartlett2002,Braunstein2005}. This capability allows Clavina to directly simulate the dynamics of an interacting many-body system, moving beyond the constraints of linear-optical networks. As a demonstration, we simulate the Bose-Hubbard model using the circuit in Fig. 4(a) \cite{Kalajdzievski2018}.

Our platform enables real-time observation of Bose-Hubbard dynamics while capable of tuning the interaction-to-tunneling ratio $U/J$ and the evolution time $t$. As demonstrations, we initialize two photons in chosen Fock states and then measure their distributions.
Fig. 4(c) demonstrates that the preferred photon-occupancy pattern depends jointly on the interaction ratio $U/J$ and the initial state. Starting from the bunched state $|200\rangle$, and fixing the evolution time $t=0.5$, weak interactions populate both bunched and separated configurations, whereas strong repulsion suppresses double occupancy and stabilizes single occupied states. The trend reverses for an initial $|110\rangle$ state. Weak interactions keep the photons apart, while increasing $U/J$ favors re-bunching into doubly occupied states. Fig. 4(d) shows the dynamics with $U = J$. For different initial states, we tracked the state populations over $t\in[0.5,3.5]$. After a long period of evolution, the dynamics show an initialization-dependent preference for either bunched or separated configurations. Further details are provided in the Supplementary Information Section VIII.

Compared with superconducting and other platforms, Clavina offers several distinct advantages. Benefiting from fast electro-optic modulation, the circuit can be reconfigured on the fly, making both the couplings $J_{ij}$ and the on-site interactions $U$ tunable in time. Such capabilities are typically unavailable in superconducting-qubit architectures \cite{Arute2019}, where $U$ can only be set as a large value and non-tunable, thereby restricting simulations to the hard-core boson limit (within a perturbative regime). On our platform, we can explore finite-$U$ physics and, by leveraging PNRDs, directly probe multi-occupancy beyond that regime. Collectively, these capabilities establish our first photonic realization of this kind of many-body model as a practical pathway to scalable photonic quantum simulators \cite{Daley2022,Aspuru2012,Lloyd1996}.

\vspace{6pt}
\noindent \textbf{Discussion and outlook}\\
To overcome the longstanding limitation of photonic quantum computing to  linear resources only, we propose an extensible architecture that integrates diverse functional modules with nonlinear resources, including either photon-number source and nonlinear gates, thus marking a crucial advance toward universality.
The measurement-induced approach used in our work circumvents the intrinsic limitations identified in Ref.~\cite{Shapiro2006} by avoiding direct Kerr nonlinearities and thus the causality-induced phase noise that compromises high-fidelity operation. This approach can also extend naturally to the multiphoton case~\cite{Fadrny2024}, enabling strong nonlinearities at higher photon numbers. For near-term implementations, alongside ongoing improvements in gate fidelity, the probabilistic nature can be mitigated by offline preparation buffered in a fiber-delay loop using a repeat-until-success strategy; the prepared gate is then injected into the main circuit via gate teleportation, enabling near-deterministic and scalable operation. Furthermore, photon addition and subtraction can be achieved deterministically via light-matter interfaces using such as semiconductor quantum dots \cite{Uppu2021}. The engineered quantum-dot devices can act as single-photon adders \cite{Sotier2009, Steindl2021} or subtractors \cite{Lund2024} on each time-bin mode, providing a nonlinear resource while preserving the platform's modularity.

In the long term, direct Kerr-like interactions may also become feasible. While Ref.~\cite{Shapiro2006} argued that single-photon Kerr nonlinearity cannot be achieved with high-fidelity under realistic conditions, recent progress indicates no fundamental barrier to such operations. Platforms like strongly coupled photonic crystal microcavities \cite{Hastrup2022}, hybrid Rydberg-atom systems \cite{Tiarks2019} and waveguide-embedded quantum dots \cite{Staunstrup2024} are steadily advancing toward significant single-photon-level nonlinearities, suggesting that high-fidelity Kerr or other nonlinear gates could eventually be realized. Importantly, these platforms are naturally compatible with Clavina's extensible time-bin loop architecture, allowing each temporal mode to interact sequentially with a shared nonlinear module for scalable, hardware-efficient nonlinear processing.

Having addressed the integration of nonlinear resources, the next challenge on the path to a large-scale practical quantum photonic computer is suppressing photon loss and other noise mechanisms. Photon-loss errors can, in principle, be corrected with the GKP code \cite{Harris2025}. In this work, we demonstrate quasi-deterministic generation of GKP states and lay the groundwork for producing high-quality GKP states with large amplitude and sufficient squeezing \cite{Winnel2024} in future work. Moreover, Clavina's temporal encoding is naturally suited \cite{Larsen2019,Asavanant2019} to generating large GKP cluster states and constructing surface codes \cite{Noh2020}, making it a promising route for realizing error correction \cite{Larsen2025,Konno2024}. Details are presented in Supplementary Information Section VII.C. In addition, Clavina is readily transferable to integrated photonic circuits---for example, those based on thin-film Lithium niobate on insulator---which are capable of fast electro-optic modulation and integrated high-gain squeezing. With continued advances in component engineering and quantum error correction, the blueprint realized in Clavina can be refined into a fault-tolerant, universal photonic quantum computer, positioning photonic systems as compelling candidates for next-generation quantum technology.

\vspace{6pt}

\noindent {\bf Acknowledgments}\\
We thank B. Q. Baragiola, Z. Yin, Z. Jia, Z. Li, M. Bellini, Z.-H. Liu, Z.-P. Zhong, A. Zhang, Z. Zhu for the helpful discussions and feedback on the manuscript. {\bf Funding:} This work was supported by UK Research and Innovation Guarantee Postdoctoral Fellowship (project: EP/Y029631/1) and Future Leaders Fellowship (project: MR/W011794/1), The Royal Society (project:  RG\textbackslash R2\textbackslash 232514), Engineering and Physical Sciences Research Council hub for Quantum Computing via Integrated and Interconnected Implementations (project: EP/Z53318X/1), Distributed Quantum Computing and Applications (project: EP/W032643/1), Schmidt Sciences, LLC., EU Horizon 2020 Marie Sklodowska-Curie Innovation Training Network (project: 956071, `AppQInfo'), National Research Council of Canada (project QSP 062-2), and the Engineering and Physical Sciences Research Council (EP/W032643/1).{\bf Author contributions:} S.Y. and R.B.P. designed the project. S.Y., R.B.P., Z.-H.Y. built and operated the experiments with the help of K.-C.C., G.Y., Z.L., E.M., G.J.M. and I.A.W. S.Y. and R.B.P. designed the cat state breeding and GKP generation circuits and completed the corresponding data analysis with the help of K.-C.C., Z.L., M.S.K., J.S. E.M., Y.K.A., G.J.M., K.-C.C., S.H.W., D.M.D.L. and Z.C.Z. J.S, S.Y. Y.D. and R.B. conceived the theoretical model and circuits of the Bose-Hubbard Hamiltonian simulation. J.S., S.Y. R.B., Y.D., V.V., M.Z., K.-C.C., S.H.W. and D.M.D.L. completed the corresponding data analysis. S.Y., J.S., R.B.P., K.-C.C., and Z.L. prepared the manuscript with input from all authors. {\bf Competing interests:} The authors declare no competing interests.

\end{document}